\documentclass[twocolumn,prl,preprintnumbers,nofootinbib]{revtex4-1}
\usepackage{graphicx, color} 
\usepackage{amsmath,amssymb}
\usepackage{ulem} 
\usepackage{epsf}
\usepackage{color}

\usepackage[italicdiff]{physics}
\usepackage{bm}
\usepackage{mathtools}
\usepackage{enumitem}
\usepackage{tensor}
\usepackage{wrapfig}
\usepackage{footnote}
\usepackage{here, comment}
\usepackage{bbm}

\usepackage{here, comment}
\usepackage{tikz-feynman}

\newcommand{\eqn}[1]{\begin{equation} #1 \end{equation}}

\newcommand{\al}[1]{\begin{align} #1\end{align}}

\allowdisplaybreaks[1]

\newcommand{\be}{\begin{equation}}
\newcommand{\e}{\end{equation}}
\newcommand{\aln}[1]{\begin{align}#1\end{align}}

\begin{document}

\preprint{KEK-TH-2307}

\title{Entanglement entropy in 
	scalar field theory \\ 
and $\mathbb{Z}_M$ gauge theory on Feynman diagrams} 
\author{Satoshi Iso$^{a,b}$} 
\email{satoshi.iso@kek.jp}
\author{Takato Mori$^{a,b}$}
\email{moritaka@post.kek.jp}
\author{Katsuta Sakai$^a$}
\email{sakaika@post.kek.jp}

\affiliation{
$^a$ KEK Theory Center, High Energy Accelerator Research Organization (KEK), Oho 1-1, Tsukuba, Ibaraki 305-0801, Japan.\\
$^b$ {The} Graduate University for Advanced Studies (SOKENDAI), Oho 1-1, Tsukuba, Ibaraki 305-0801, Japan.
}

\begin{abstract} 
Entanglement entropy (EE) in interacting field theories has two important issues: 
renormalization of UV divergences and non-Gaussianity of the vacuum.
In this paper, we investigate them in the framework of the two-particle irreducible formalism.
In particular, we consider EE of a half space in an interacting scalar field theory. 
It is formulated as $\mathbb{Z}_M$ gauge theory on Feynman diagrams: 
$\mathbb{Z}_M$ fluxes are assigned on plaquettes and summed to obtain EE. 
Some configurations of fluxes are interpreted as twists of propagators and vertices. 
The former gives a Gaussian part of EE written in terms of a renormalized 2-point function 
while the latter reflects non-Gaussianity  of the vacuum. 
\end{abstract}

\maketitle


\section{I. Introduction}
Entanglement entropy (EE) provides important information of a given 
state, in particular,
correlations in a ground state between two spatially separated regions and 
has been widely discussed in quantum information, condensed matter physics and, even in quantum gravity, {cosmology,} and high energy physics \cite{PhysRevA.66.032110,Osterloh_2002,PhysRevLett.90.227902,Jin_2004,Calabrese_2004,Ryu:2006bv,Ryu:2006ef,Hubeny:2007xt,Almheiri:2020cfm,Almheiri:2019hni,Penington:2019kki,Nambu:2008my}.
Despite its  importance, the practical computation of EE in field theories is not an easy task 
 and consequently, much of the works  have focused on Gaussian states \cite{PhysRevA.70.052329,Katsinis:2017qzh,Bianchi:2019pvv,Solodukhin:2011gn,Lewkowycz:2012qr,Hertzberg:2010uv,Buividovich:2018scl}, low-energy sectors of conformal field theories (CFTs) \cite{Calabrese_2004,Ruggiero:2018hyl,Hung:2014npa,Casini:2010kt} or holographic CFTs \cite{Ryu:2006ef,Ryu:2006bv,Nishioka_2009}.
 {For Gaussian states, both the so-called real-time approach and the imaginary time approach as known as the replica trick \cite{Casini_2009} are applicable. The computations make use of its Gaussianity; the reduced density matrix is still Gaussian \cite{Chen:2020ild}. For CFTs, EE of quite general shapes of subregions can be studied while the conformal symmetry plays an important role in reducing the problem simpler and tractable. For theories with holographic duals, EE can be computed in a simple, easy manner as a geometric quantity while the existence of the AdS/CFT correspondence \cite{Maldacena:1997re,Gubser:1998bc,Witten:1998qj} is obviously necessary.}
Many features of EE were clarified, but 
there are only a few studies on EE in interacting theories: a perturbation from free theories \cite{Hertzberg:2012mn,Chen:2020ild} or CFTs \cite{Rosenhaus:2014zza}, {the renormalization group flow given fixed point CFTs \cite{Whitsitt:2016irx},} and large $N$ expansions \cite{PhysRevB.80.115122,Akers:2015bgh}. 
{There are also some nonperturbative studies \cite{Rosenhaus:2014ula,Cotler:2015zda,Fernandez-Melgarejo:2020utg,Herzog:2013py,Wang_2014,Buividovich:2008kq,Buividovich:2008gq,Itou:2015cyu,Rabenstein:2018bri}, 
but their analytical evaluations are difficult.}
Our goal in this paper is to provide a {field theoretic,} systematic way to explore EE in {a} massive interacting theory, {which is neither free nor conformally invariant and the existence of its holographic dual is not assumed}.

{Besides computability, EE has an obvious problem specific to field theories.}
Since field theories contain infinitely many degrees of freedom, EE suffers from ultraviolet divergences and 
an appropriate regularization and renormalization are necessary to obtain  finite results. 
{For free theories, the UV-divergent EE can be regularized by suitably renormalizing parameters in the background gravity \cite{Cooperman:2013iqr,Barrella:2013wja,Taylor:2016aoi,Taylor:2020uwf,Liu:2012eea,Liu:2013una}. 
There are  additional UV divergences  in interacting field theories, which
should be dealt with  the usual flat space renormalization. 
A perturbative treatment of 
this renormalization was discussed  \cite{Hertzberg:2012mn}.

In this paper, we give a systematic study of EE in interacting field theories.
We consider a scalar field theory with $\phi^4$ interactions in a simple geometrical setup, a half space being traced over.
It is formulated as a $\mathbb{Z}_M$ gauge theory on Feynman diagrams: 
We perturbatively evaluate EE in the two-particle irreducible (2PI) formalism 
and obtain a generalized 1-loop type expression of EE in terms of renormalized propagators. 
Moreover, we show that the non-Gaussian nature of the vacuum wave function gives further corrections
to EE associated with 4-point vertex functions.

For the Hilbert space 
{composed of} two subsystems {on a time slice,} 
$\mathcal{H}_{\mathrm{tot}}=\mathcal{H}_A\otimes\mathcal{H}_{\bar{A}}$, 
the EE for $A$ is defined as 
$ S_{A}=-\Tr_{A}\rho_{A}\log\rho_{A}$,
where $\rho_A=\Tr_{\bar{A}}\rho_\mathrm{tot}$ is a reduced density matrix of 
the total one, $\rho_\mathrm{tot}$. 
In this paper, we choose the subregion $A$ as a half space specified by $A=\{x^0=0,x_\perp\ge 0, \forall x_\parallel\}$
and $\bar{A}$ as its complement, 
where $x_\perp$ and $x_\parallel$ are the normal and parallel directions to $\partial A$ respectively.
A standard method to calculate EE $S_A$ is {known as} the replica method \cite{Holzhey:1994we,Calabrese_2004}, where
$ S_A  =\lim_{n\rightarrow 1}\left[(1/(1-n)\log \Tr\rho_A^n \right].$
{Let us define an unnormalized density matrix $\tilde{\rho}_\mathrm{tot}$ by $\rho_\mathrm{tot}=\tilde{\rho}_\mathrm{tot}/Z_1$, where $Z_1$ is a partition function on $\mathbb{R}^{d+1}$ as a Euclidean path integral.
Then, using $\tilde{\rho}_A = \Tr_{A} \tilde{\rho}_\mathrm{tot}$, an unnormalized reduced density matrix, $\Tr \tilde{\rho}_A^n$ can be viewed as
}
a partition function on $\Sigma_n\times\mathbb{R}^{d-1}$, where 
$\Sigma_n$ is an $n$-folded cover of a two-dimensional plane and, thus, a 
two-dimensional cone with deficit angle $2\pi(1-n)$.
The EE can be rewritten in terms of the free energy 
$F_n\equiv -\log Z_n \equiv -\log \Tr_A\tilde{\rho}_A^{\,n}$ as 
$S_A=  \partial F_n/ \partial n |_{n\to1}-F_1$.

\section{II. Area law of EE in orbifold method}
We first show the area law of EE. 
For this purpose,  the orbifolding method \cite{Nishioka_2007,He:2014gva} is convenient. We consider a space $\mathbb{R}^2/\mathbb{Z}_M$
instead of the $n$-folded space, $\Sigma_n$. Since $M$ can be interpreted as $n=1/M$,
the {vacuum} EE on the $\mathbb{Z}_M$ orbifold is given in terms of the free energy on the $\mathbb{Z}_M$ orbifold $F^{(M)}=F_{1/n}$ as
\aln{
S_A=-\frac{\partial \left(M F^{(M)}\right) }{\partial M}\bigg|_{M\to 1},
\label{eq:EE_M}
}
{provided $M\in\mathbb{Z}_{>1}$ can be analytically continued to 1. }A state on the orbifold can be obtained by acting the $\mathbb{Z}_M$ projection operator, 
$
\hat{P}=\sum_{n=0}^{M-1}\hat{g}^{\, n} /M
$
on a state in an ordinary two-dimensional plane, where $\hat{g}$ is a $2\pi/M$ rotation operator {around the origin}. 
{
In this paper, we call a $\mathbb{Z}_M$ rotation $\hat{g}^n$ as an $n\,(\in \mathbb{Z} \mod M)$ \it{twist} operation.}

Let us consider, for simplicity, a scalar field theory on the $\mathbb{Z}_M$ orbifold without a nonminimal coupling
to the curvature. {Since scalar fields have no spin and are singlet under the spatial rotation, the $\mathbb{Z}_M$ action $\hat{g}$ on the internal space of the fields is trivial. Its explicit action is given as follows: 
\eqn{
\hat{g} [\phi (x)]=\phi (\hat{g} x)=\phi (\hat{g}\bm{x},x_\parallel)=\phi( e^{2\pi i/M} x_\perp, e^{-2\pi i/M} \bar{x}_\perp, x_\parallel),
}
where the two-dimensional coordinates on $\mathbb{Z}_M$ is given by $\bm{x}$ or equivalently by the complex coordinates $(x_\perp,\bar{x}_\perp)$. The remaining codimension-two coordinates parallel to the subregion boundary are given by $x_\parallel$. The total $(d+1)$-dimensional coordinates are denoted by $x$.} 
The action {for the scalar field theory on $\mathbb{Z}_M$ orbifold} is given
in terms of the field $\phi(x)$ on {a flat space} $\mathbb{R}^2\times\mathbb{R}^{d-1}$ as 
\aln{
\int \frac{d^2x}{M} d^{d-1}x_\parallel \left[\frac{1}{2}\phi \hat{P} \left(-\Box+m^2 \right)\hat{P}\phi+V(\hat{P}\phi)\right]. 
\label{eq:action} 
}
In the following, we consider $\lambda \phi^4/4$ potential {for simplicity}. {However, this particular choice of the potential is only for simplicity and generalizations to the other form of potentials such as cubic or higher orders are straightforward.} 
{From the action Eq.\eqref{eq:action}, the inverse propagator can be read off as
\[
\hat{G}_0^{-1\, (M)}=\hat{P}\hat{G}_0^{-1}\hat{P}.
\]
Since the propagator is its inverse on the $\mathbb{Z}_M$ orbifold, it satisfies the relation
\[
\hat{G}_0^{-1\, (M)} \frac{1}{M} \hat{G}_0^{(M)}=\hat{G}_0^{(M)} \frac{1}{M} \hat{G}_0^{-1\, (M)}=\hat{P}.
\]
Thus,} the propagator on the orbifold is written as
\aln{
G_{0}^{(M)} (x,y) &= M\bra{x}(\hat{P} {\hat{G}_0} \hat{P})\ket{y} 
=\sum_{n=0}^{M-1} G_0(\hat{g}^n x,y) ,
\label{eq:green}
}
where 
\aln{ 
G_0(\hat{g}^n x, y) = \bra{\hat{g}^n x} {\hat{G}_0} \ket{y} =  \int\frac{d^{d+1} p}{(2\pi)^{d+1}}\frac{e^{i p \cdot(\hat{g}^n x-y)}}{p^2+m^2} .
}
The projection operator on $y$ is eliminated by a rotation of the momentum $p$.
Since $p\cdot \hat{g}^n x=\hat{g}^{-n} p \cdot x$, we see that
 the flow-in momentum from the propagator at a vertex $x$ is given by the twisted momentum, $\hat{g}^{-n} p$. {Twists of coordinates are equivalent to the inverse twists of the corresponding momenta.}
 
{For the calculation of EE, we need to compute the free energy, which is minus the sum of the all possible connected bubble diagrams.} Consider a Feynman diagram with $N_V$ vertices, $N_P$ propagators, and $L$ loops. 
At each vertex, there is a factor 
{
\[
{-}\frac{\lambda}{M},
\]
where $1/M$ comes from the integration measure in Eq.\eqref{eq:action}.
 Thus,
}
an overall $M$ dependence seems to be given by $(1/M)^{N_V}$.
But it is not correct since $N_V-1$ of the projection operators in the $N_P$
propagators can be further eliminated by rotations 
of coordinates at the vertices.
It can be understood as follows.
If we particularly pay attention to a propagator $G_0(\hat{g}^n x, y)$ and a vertex $x$, 
the twist $\hat{g}^n$ in the propagator can be eliminated by
changing the integration variable $x$.
Thus the summation of the twist $(n =0, \cdots M-1 )$ eliminates the $1/M$ factor at the vertex. 
This procedure can be continued only up to $N_V-1$ vertices. 
The last integration of the coordinates of a vertex cannot absorb a twist of propagators.
In ordinary flat space without twists, due to the translational invariance, 
the integration gives the volume of the space-time, $(2\pi)^{d}\delta^d(0)=V_d$. 
In our case with twists, reflecting the absence of the translational invariance on the orbifold, 
the last $x$-integration instead gives
$V_{d-1}  \delta^2 (\sum_{l=1}^L (1-\hat{g}^{-n_l}) \bm{k}_l)$. {This procedure of eliminating redundant twists is depicted in Fig.\ref{fig:phi43loops} , giving a 3-loop bubble diagram as an example.
On the right figure, a twist of the propagator (the bottom dashed line) in the left figure is removed by 
a rotation at a vertex (either left or right point). Accordingly the coefficient $M$ appears from the sum of all the twists
from 0 to $M-1$.
}
\begin{figure}
	\begin{tabular}{c}
		\hspace*{-0.05\linewidth}
		\begin{minipage}{0.5\hsize}
			\centering
			\includegraphics[width=\linewidth]{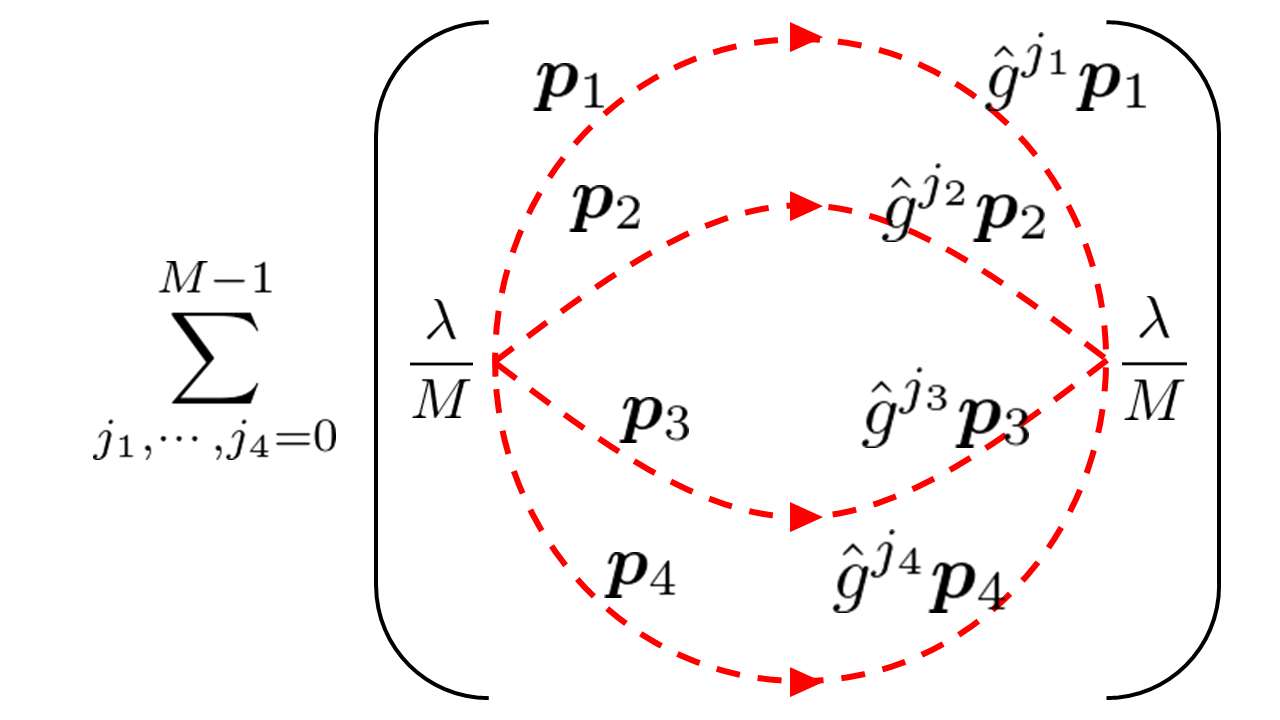}\label{fig:bubble1}
		\end{minipage}
		\hspace*{0.01\linewidth}
		\begin{minipage}{0.5\hsize}
			\centering
			\includegraphics[width=\linewidth]{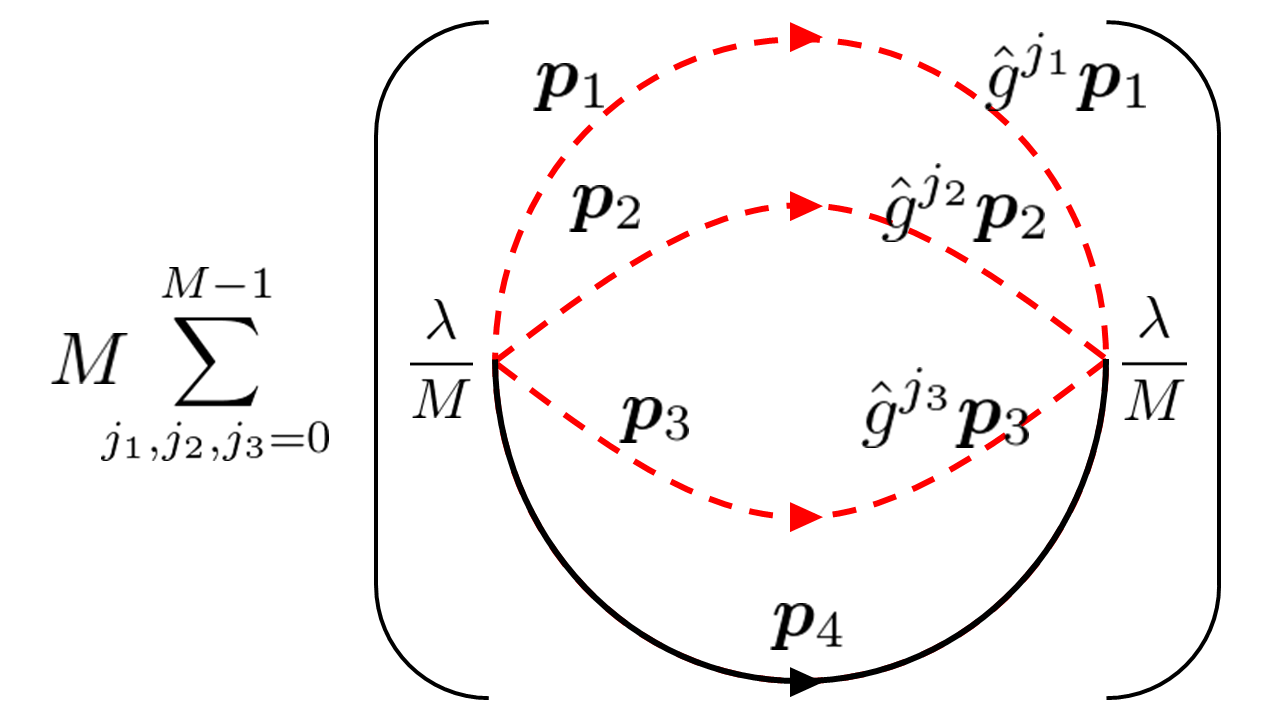}\label{fig:bubble2}
		\end{minipage}
	\end{tabular}
	\caption{{Two equivalent sets of three-loops diagrams in the $\phi^4$ theory on the orbifold. Red dashed lines denote the propagators with their momenta twisted. A black line in the right figure represents the ordinary propagator in the flat space.}}
	\label{fig:phi43loops}
\end{figure}

Now the area law of EE is proved as follows.
After eliminating the redundancies of twists as above, there are $L=N_P-N_V+1$ nontrivial twists and
the overall $M$-dependence of the free energy 
$F^{(M)}$ is given by $1/M$. 
The ordinary volume factor $(2\pi)^{d+1}\delta^{d+1}(0)=V_{d+1}$ in $F^{(M=1)}$
is being replaced by the area $V_{d-1}$ of the boundary of the subregion  times 
an additional factor $\delta^2 (\sum_{l=1}^L (1-\hat{g}^{-n_l}) \bm{k}_l)$ in the momentum integrations. 
Note that the additional factor gives the two-dimensional volume $V_2$ only when all $n_l=0$.
Due to the overall $1/M$ factor, 
$V_{d+1}$-proportional terms in $F^{(M)}$, i.e., all $n_l=0$ ($l=1 \cdots L$), 
are canceled in Eq.(\ref{eq:EE_M}),
and do not contribute to EE, while the other terms, such that some of $\{n_l\}$ are nonvanishing, are proportional 
to the area $V_{d-1}$ and contribute to EE. 
This analysis holds to all orders in the perturbation theory.

\section{III. $\mathbb{Z}_M$ gauge theory on Feynman diagrams}
The orbifold field theory can be regarded as $\mathbb{Z}_M$ gauge theory
on Feynman diagrams. 
\begin{figure}[h]
	\begin{tabular}{c}
		\hspace*{-0.05\linewidth}
		\begin{minipage}{0.35\hsize}
			\centering
			\includegraphics[width=\linewidth]{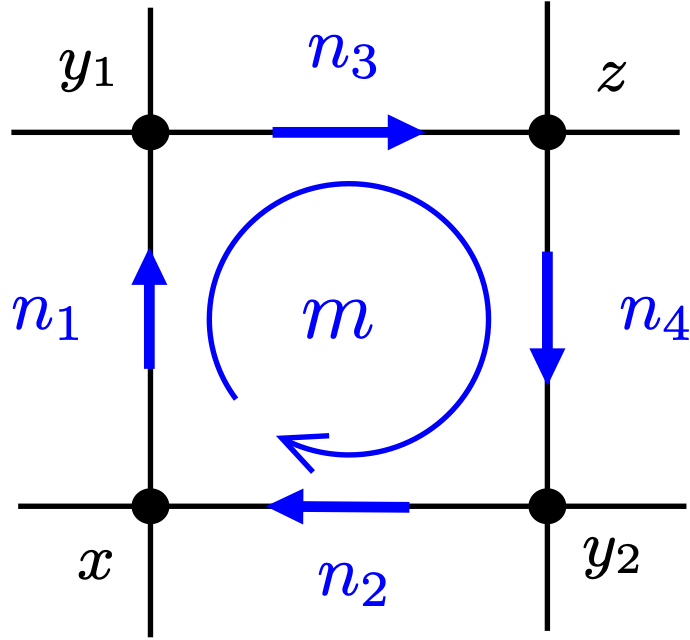}
		\end{minipage}
	\hspace*{0.05\linewidth}
		\begin{minipage}{0.45\hsize}
			\centering
			\includegraphics[width=\linewidth]{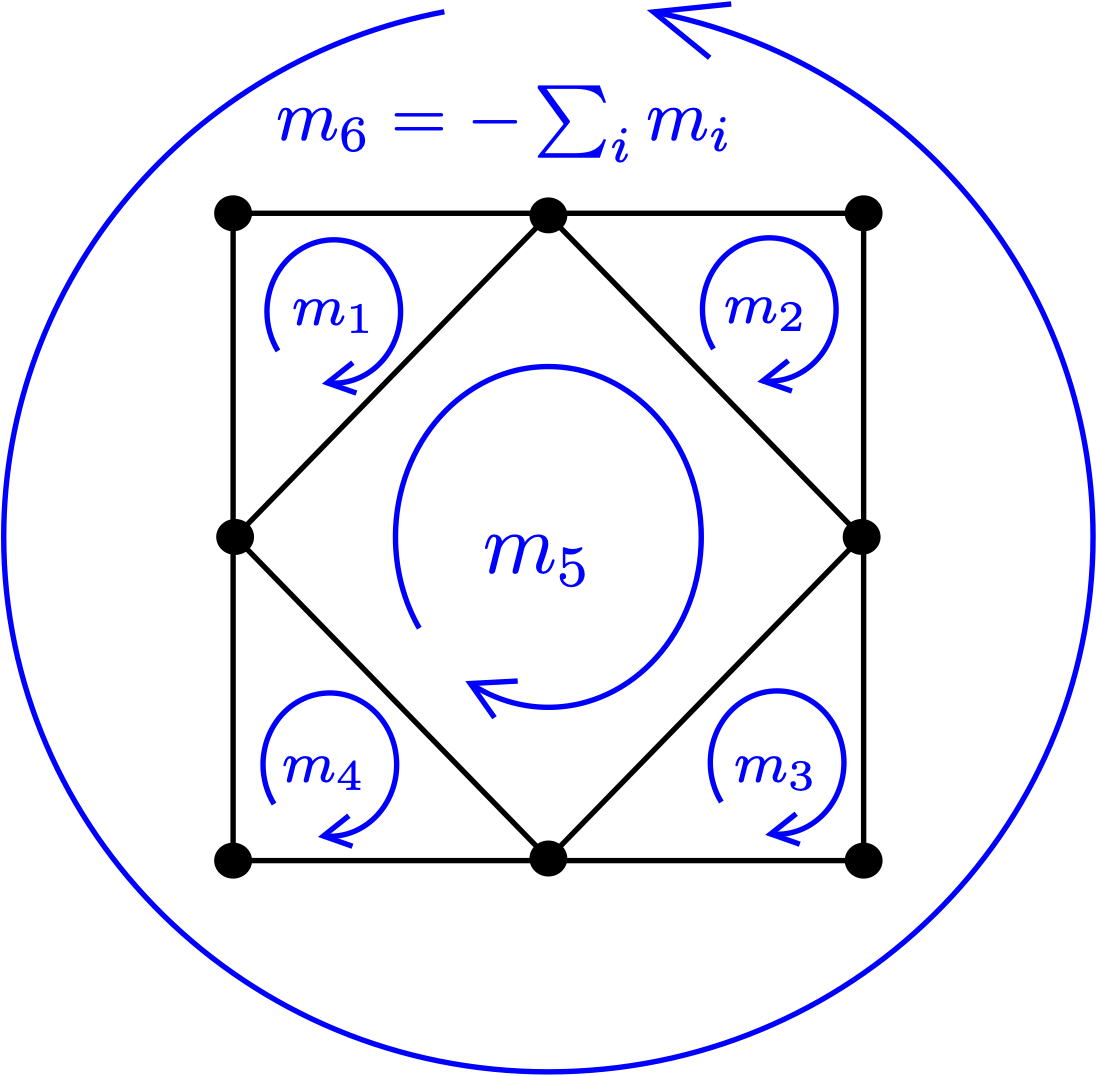}
		\end{minipage}
	\end{tabular}
\caption{
$\mathbb{Z}_M$ gauge theory on Feynman diagrams: $\{n_i\}$ are twists on links (propagators), and $m=\sum_i n_i$ is 
a flux of twists around the plaquette and invariant under  $\mathbb{Z}_M$ gauge transformations on vertices.
The right figure is a set of $\mathbb{Z}_M$ invariant fluxes of twists on plaquettes.}
\label{Fig1}
\end{figure}
On a $\mathbb{Z}_M$ orbifold, each propagator in a Feynman diagram is twisted as in Eq.(\ref{eq:green}). 
A rotation of the coordinates at the vertex $x$ in Fig.\ref{Fig1} by $2\pi l/M$ shifts $n_1$ by $l$, and $n_2$ by $-l$; 
therefore, {the sum} of twists around a plaquette $m=\sum_i m_i$, {which we call a \it{flux},} is invariant under $\mathbb{Z}_M$ rotations at vertices.
Consequently, for a given Feynman diagram such as the right figure of Fig.\ref{Fig1}, 
a $\mathbb{Z}_M$ invariant set of twists is given by a set of $L (=5)$ fluxes of twists on each plaquette of the Feynman diagram.
The twist of the outer circle is given by $m_6=- \sum_{i=1}^5 m_i$, since the direction of the twist is opposite
when the diagram is put on a sphere. 
We can calculate a contribution to EE from a Feynman diagram in the following procedure:
 (1) assign twists $m_i$ (mod $M$) to each plaquette of the diagram, (2) perform momentum integrations
 and evaluate free energy of the Feynman diagram for a configuration of twists $\{m_i \}$, (3) sum over all the twists $\{m_i\}$.
As discussed in the previous section, when all the twists are trivial, i.e., $m_i=0$, it does not 
contribute to EE since the overall factor $1/M$ of the free energy is canceled in Eq.(\ref{eq:EE_M}). 
Thus, we are interested in a configuration of twists, in which some of them are nonvanishing.

Let us begin with a 1-loop diagram. In the following, we write $(d+1)$-dimensional momenta and coordinates
as $(\bm{k}, k_\parallel)$ and $(\bm{x}, x_\parallel)$.
For a 1-loop diagram, there is a single twist $n$ {(Fig.\ref{Fig2})}.
\begin{figure}[h]
\centering
\includegraphics[width=0.25\linewidth]{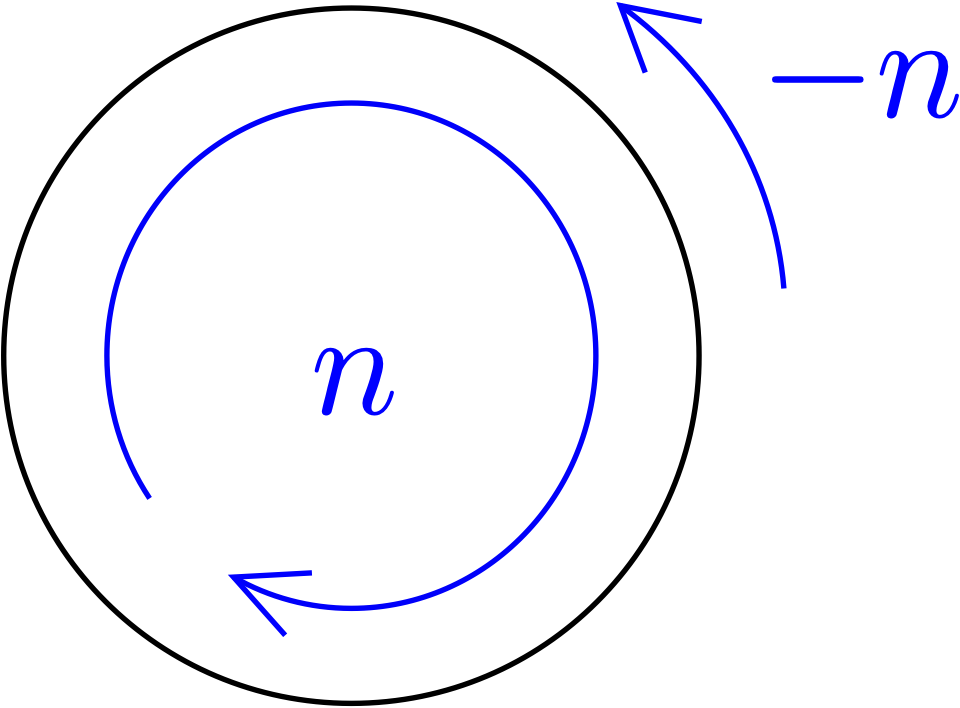}
\caption{
{There is a single twist for 1-loop diagram.}
}
\label{Fig2}
\end{figure}
The free energy with twist $n$ is easily calculated \cite{Nishioka_2007} by noting 
that $\bra{\bm{k}} \hat{g}^n \ket{\bm{k}}= (2\pi)^2 \delta^2 (\bm{k}) /4 \sin^2(n \pi /M)$ for $n \neq 0$. 
Thus, we have
\aln{
F_{\text{1-loop}}^{(M)} &
{
	= \frac{1}{2}\Tr \left[\log G_0^{-1}\right]^{(M)} 
}
\nonumber\\
&
{
	= \frac{V_{d-1}}{2} \int \frac{\dd[d+1]{k}} {(2 \pi)^{d+1}} \log (k^2+m^2) \ev{\hat{P}}{k}
}
\nonumber\\
&= \frac{V_{d-1} }{2 M} \int \frac{d^2 \bm{k} \ d^{d-1}k_{\parallel}} {(2 \pi)^{d-1}}   \log (k ^2 + m^2)
\nonumber \\
& \times \left( \frac{V_2}{(2\pi)^2}+  \sum_{n=1}^{M-1} \frac{\delta^2(\bm{k})}{4 \sin^2(\frac{n \pi}{M})} \right) .
}
The volume factor proportional to $V_2$ vanishes in Eq.(\ref{eq:EE_M}). By using the relation
$\sum_{n=1}^{M-1} 1/\sin^2(n\pi/M)=(M^2-1)/3$, we obtain the EE
\aln{
S_{\text{1-loop}}
&{
	=-\left.\pdv{M}\left[M F_{\text{1-loop}}^{(M)}\right]\right\vert_{M\rightarrow 1}
}
\nonumber\\
&=- \frac{V_{d-1} }{12}  \int^{1/\epsilon} \frac{d^{d-1}k_{\parallel}} {(2 \pi)^{d-1}}   
\log \left[ (k_{\parallel} ^2 + m^2) \epsilon^2 \right].
\label{EE-1loop}
}
{Here a UV cutoff scale $\epsilon$ is introduced. Note that EE decreases as the mass increases.}
The appearance of the area law can be interpreted as pinning of the propagator $G_0 (x,y)$ at the origin of the orbifold
as demonstrated below. A twisted propagator of $G_0(x, y) :=G_0 (\bm{r}; r_\parallel)$ is written as
\aln{
& G_{0} (\hat{g}^n x -y)  =  
G_{0} ( \hat{g}^{n/2} \bm{x} - \hat{g}^{-n/2} \bm{y} ; r_\parallel)  \nonumber \\ &
= G_0 (\cos \theta_n r_i + 2 \sin \theta_n \epsilon_{ki} X_k ; r_\parallel) 
\nonumber \\
&= e^{\cot \theta_n \hat{R}_X /2} G_0 (2 \sin \theta_n  \bm{X} ;  r_\parallel),
\label{twistedpropagator}
}
where 
 $\hat{R}_X = \epsilon_{ij} r_i \partial_{X_j}$, $r=x -y$, $\bm{X}=(\bm{x}+\bm{y})/2$ and $\theta_n= n\pi /M$.
Suppose that the twisted propagator is multiplied by a function $F(r)$ of the relative coordinate $r$ and integrated as
$I= \int dx  dy \ G_{0} ( \hat{g}^n x- y) F(r)$. Such integration appears when there are no more twists in the Feynman diagram. 
Then, due to 
$\hat{R}_X F(r)=0$, the twisted propagator $G_{0} (\hat{g}^n x -y)$ can be replaced by $G_0 (2 \sin\theta_n  \bm{X} ;  r_\parallel)$.
For $n \neq 0$, by rescaling momentum $\bm{p}$,  it is written as 
\aln{
& G_0 \left(2 s_n \bm{X}  ;  r_\parallel \right)  
= 
\frac{1}{4 s^2_n} 
  \int \frac{d^2 \bm{k} d^{d-1} k_\parallel}{(2\pi)^{d+1}} \frac{e^{i  \bm{k} \cdot \bm{X} + i k_\parallel \cdot r_\parallel}}
{\left(\bm{k}^2/4 s^2_n \right) + M^2_{k_\parallel}} \nonumber \\
&=\frac{1}{4 s^2_n}  
  \int \frac{ d^{d-1} k_\parallel}{(2\pi)^{d-1}}
{
	e^{ i k_\parallel \cdot r_\parallel}
	\frac{1}
	{\left(-\partial^2_X/4 s^2_n \right) + M^2_{k_\parallel}} 
}
\delta^2 (\bm{X}) ,
\label{G0expand}
}
where $M^2_{k_\parallel} :=k_\parallel^2+m^2 $ and $s_n :=\sin \theta_n$.
{Since $\partial_X$ is set to zero via integration by parts in the $\bm{X}$ integration,}
the coordinate $\bm{X}=(\bm{x}+\bm{y})/2$ is pinned at the origin of the orbifold. 
It is straightforward to see that 
$S_{\text{1-loop}}$ in Eq. (\ref{EE-1loop}) can be reproduced by using this pinned propagator.
Note that, when there is another twist in the Feynman diagram, the function $F$ depends 
on $\bm{X}$ and derivative terms in Eq. (\ref{twistedpropagator}) cannot be dropped. 

Next let us consider a figure-eight 2-loop diagram of Fig.\ref{Fig3} with  twists $(m_1, m_2)$.
Its free energy is given by 
\aln{
F_{\text{2-loop}}^{(M)}=
\sum_{m_1,m_2} \frac{3 \lambda}{4M} \int d^{d+1} x \ G_0 (\hat{g}^{m_1} x, x) G_0(\hat{g}^{ m_2} x, x) .
\label{8figure}
}
\begin{figure}[h]
\centering
\includegraphics[width=0.5\linewidth]{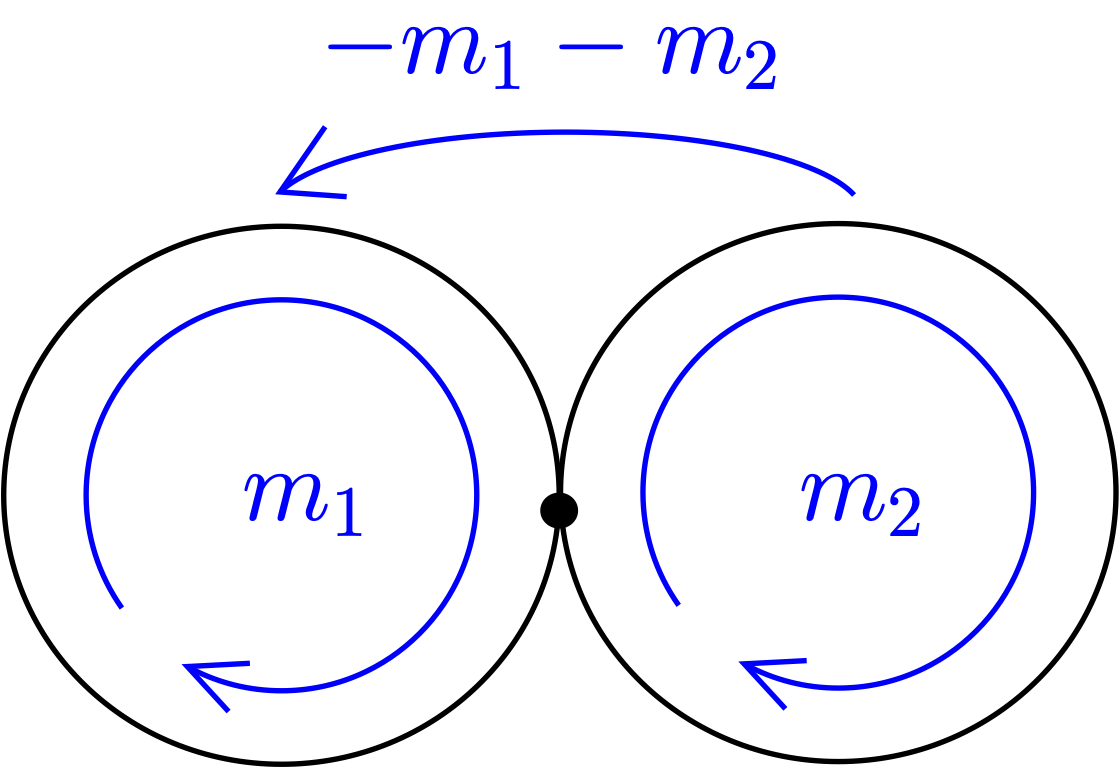}
\caption{
A 2-loop diagram with twist $(m_1, m_2)$. 
}
\label{Fig3}
\end{figure}
Specific configurations of twists, $(m \neq 0, 0)$ and $(0, m \neq 0)$, correspond to a twist of each propagator {(Fig.\ref{Fig3-1})} and 
renormalize the mass of the bare propagator in Eq.(\ref{EE-1loop}) \cite{Hertzberg:2012mn}. 
The corresponding EE {to first order in $\lambda$} is given by
{
\al{
	S_\mathrm{2-loop}^\mathrm{propag.} =-\frac{V_{d-1}}{12} G_0 (0) (3\lambda G_0(0)).
}
This is nothing but $S_{\text{1-loop}}$ of Eq.(\ref{EE-1loop}) with the mass 
}
replaced by  $m^2+\delta m^2$, where $\delta m^2 = {3} \lambda G_0(0) $. 
Renormalization of propagators is one important aspect of EE in interacting field theories. 
\begin{figure}[h]
	\centering
	\includegraphics[width=\linewidth]{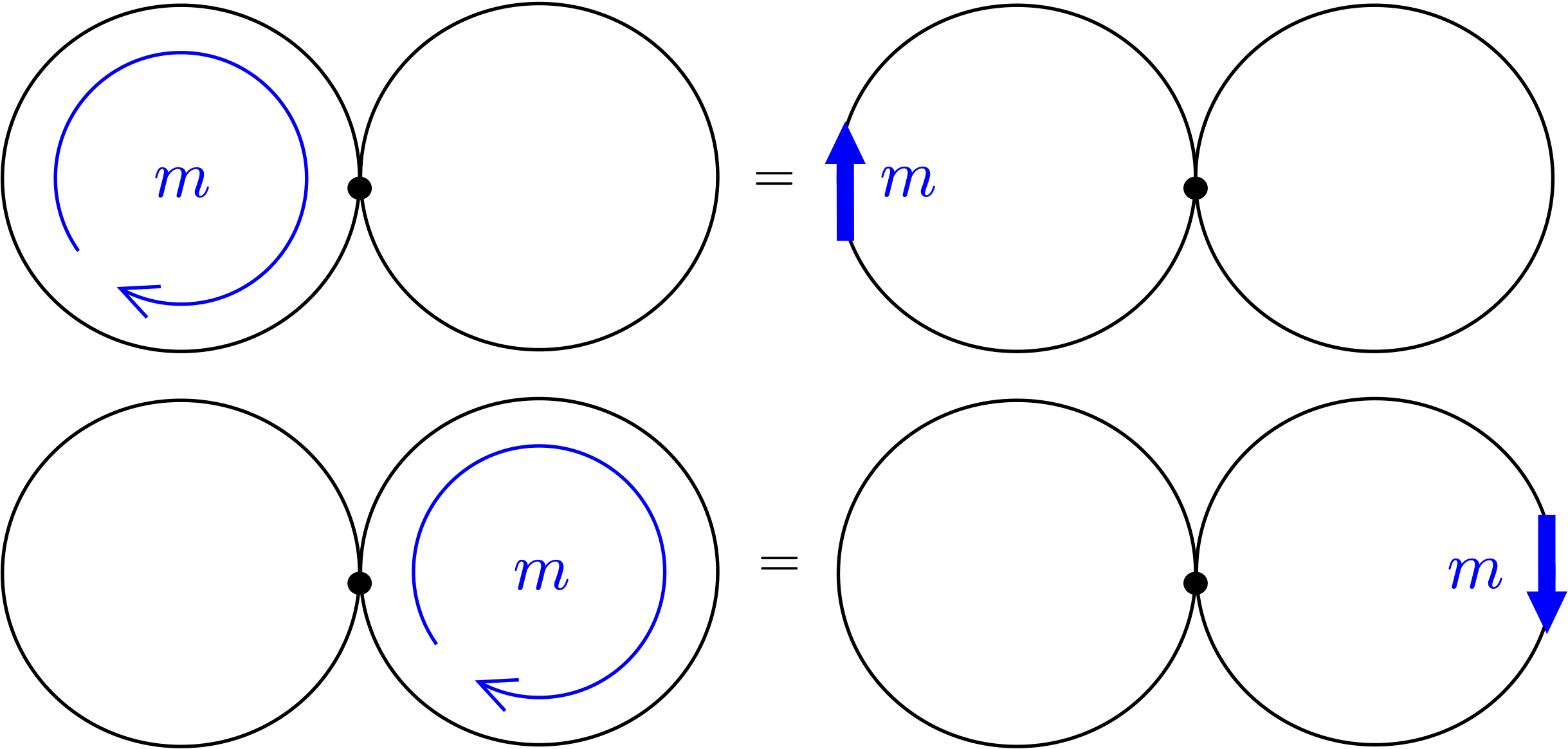}
	\caption{
		{
			2-loop diagrams with a single twist $(m_1, m_2)=(m,0),(0,m)$.These single flux can be interpreted as a twist of each propagator.
		}
	}
	\label{Fig3-1}
\end{figure}

There is another nontrivial contribution to EE from the twists $(m, \pm m)$ in Eq.(\ref{8figure}), 
which is interpreted as 
twisting the 4-point vertex {(Fig.\ref{Fig3-2})}. By rewriting the integral of Eq.(\ref{8figure}), for $m_2=-m_1$, as
\aln{
 &\int d^{d+1} x d^{d+1} y \ G_0(\hat{g}^{m_1} x, y) G_0(\hat{g}^{-m_1} y, x) \delta^{d+1}(x-y)
 \nonumber \\
 &= \int d^{d+1} x d^{d+1} y \ (G_0(x,y))^2  \ \delta^{d+1}(\hat{g}^{-m_1}x - y).
\label{8figure-2}
}
\begin{figure}[H]
	\centering
	\includegraphics[width=\linewidth]{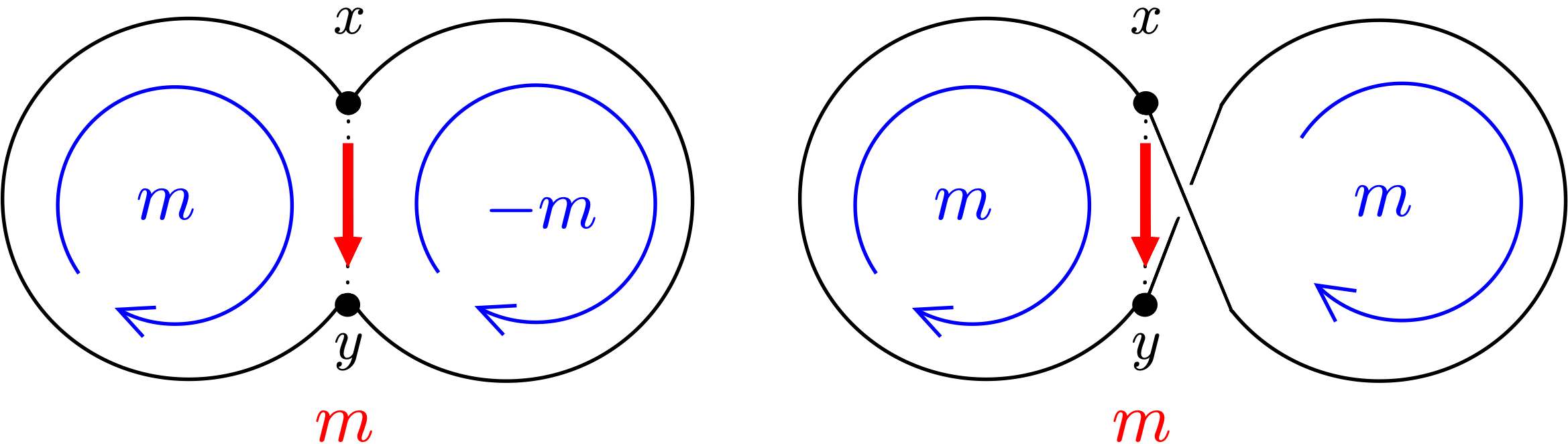}
	\caption{
		{
		2-loop diagrams with twist $(m_1, m_2)=(m,\mp m)$. These simultaneous twists from two fluxes can be interpreted 
		as a twist of the 4-point vertex by decomposing it into two 3-point vertices. 
		}
	}
	\label{Fig3-2}
\end{figure}
The same interpretation follows for $m_2=m_1$ 
{as
\aln{
	&\int d^{d+1} x d^{d+1} y \ G_0(\hat{g}^{m_1} x, y) G_0(\hat{g}^{m_1} x, y) \delta^{d+1}(x-y)
	\nonumber \\
	&= \int d^{d+1} x d^{d+1} y \ (G_0(x,y))^2  \ \delta^{d+1}(\hat{g}^{-m_1}x - y).
	\label{8figure-3}
}
}
From Eqs.(\ref{twistedpropagator}) and (\ref{G0expand}),  
we can replace 
\aln{
\delta^{2}(\hat{g}^{n}\bm{x} - \bm{y}) =
e^{\cot \theta_n \hat{R}_X /2} \frac{ \delta^2 (\bm{X})}{4 s_n^2}  \rightarrow \frac{\delta^2 (\bm{X})}{4 s_n^2} 
}
in the integral. Hence, the effect of twisting is interpreted as pinning of the position of the vertex at the origin.
{By the above replacements, we obtain the 2-loop contribution from twisting the vertex in the free energy
\eqn{F_{\mathrm{2-loop}}^{\mathrm{vertex}} =2 \frac{3\lambda}{4M} V_{d-1} \frac{M^2-1}{12} \int d^{d+1} r  (G_0(r))^2 \  \delta^{d-1}(r_\parallel).}
}
The 2-loop vertex correction to EE is then given by 
\aln{
S_{\text{2-loop}}^{\text{vertex}} 
& =  - {\frac{1}{4}}
V_{d-1} \lambda \int d^{d+1} r  (G_0(r))^2 \  \delta^{d-1}(r_\parallel).
\label{EE-2loop-vertex}
}
{The vertex correction to EE is negative for repulsive (positive $\lambda$) interaction. }
In contrast to the twisting of  propagators, 
it essentially originates from the non-Gaussianity of the vacuum. 
We also emphasize the importance of interpreting twisting in terms of 
$\mathbb{Z}_M$ fluxes on plaquette. 
If we took a special gauge and assigned twists on  particular links of Feynman diagrams, 
we could not find vertex corrections to EE since they are hidden in twisting multiple links.

Now we wonder what contributions to EE come from the other twists of the figure-eight diagram; {Fig.\ref{Fig3} with $m_1$ and $m_2$ both nonzero and $(m_1,m_2)\neq (m,\pm m)$}. Performing the integration of Eq. (\ref{8figure}), we have
 \aln{
&  \int d^{d+1} x \ G_0(\hat{g}^{m_1} x, x) G_0(\hat{g}^{m_2} x, x)     \nonumber \\
& = 
{
\frac{V_{d-1}  }{16 \pi}
 \int \frac{d^{d-1}k_\parallel d^{d-1}p_\parallel}{(2\pi)^{2(d-1)}}
	\frac{\log\left(s_{m_1} ^2 M_{k_\parallel}^2 / s_{m_2} ^2 M_{p_\parallel}^2  \right)}
	    { s_{m_1} ^2 M_{k_\parallel}^2 - s_{m_2} ^2 M_{p_\parallel}^2 } .
	    }
\label{8fig-twist}
}
\begin{figure}[h]
\centering
\includegraphics[width=\linewidth]{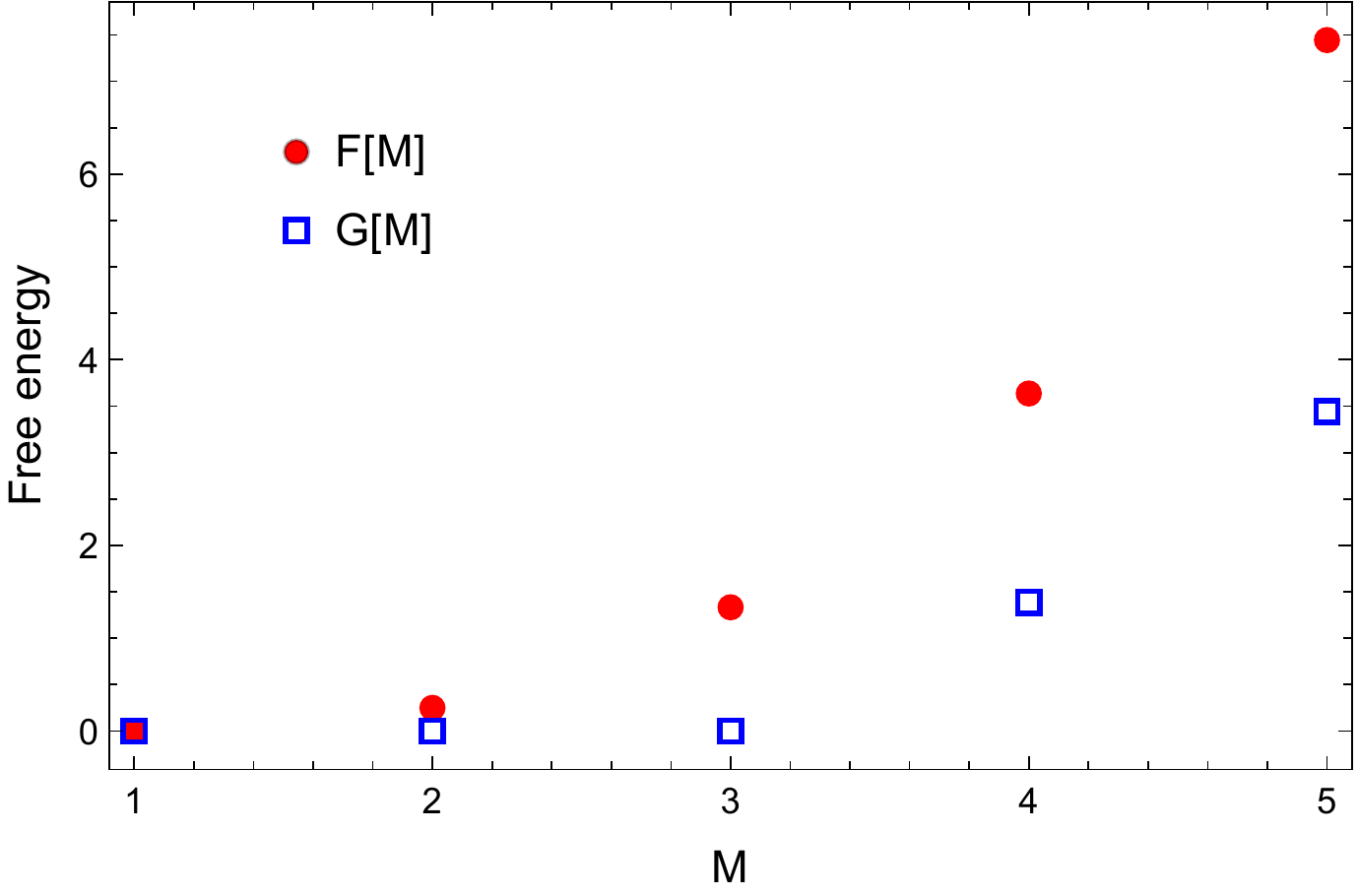} %
\caption{
$F(M)$ is a sum of the 
{integrand} of Eq.(\ref{8fig-twist}) over  $m_1, m_2 = 1 \cdots M-1$ {for $d=1$}.
 Vertex contributions  $(m, \pm m)$ are subtracted in $G(M)$.
}
\label{8fig}
\end{figure}
EE is obtained by the analytical continuation of $M$ and calculating the coefficient of the first derivative at $M=1$. 
To see the behavior of $M$-dependence of Eq.\eqref{8fig-twist}, let us focus on the $d=1$ case for simplicity.
Summation over {nonzero} $m_1$ and $m_2$ can be 
explicitly evaluated and plotted in Fig.\ref{8fig}. 
$F(M)$ in Fig.\ref{8fig} is a sum of the integrand of Eq. (\ref{8fig-twist}) over 
$m_1, m_2 =1, \cdots, M-1$. They include 2-loop vertex corrections {$(m_1,m_2)=(m,\pm m)$}. 
$G(M)$ is plotted without the vertex corrections. 
If we can simply interpolate the free energy to continuous $M$ near $M=1$, 
the first derivative seems to be dominated by the vertex contributions.  
Of course, it is not sufficient but we expect that 
EE of the figure-eight diagram is
dominantly given by twisting the propagators, $(m,0)$ and $(0,m)$, 
and the vertex $(m, \pm m)$.

\section{IV. EE in 2PI formalism}
To study the renormalization of propagators systematically, 
we calculate EE in interacting field theories in the framework of the 2PI formalism \cite{PhysRevD.10.2428,Berges:2004yj}.
The 2PI effective action is given, in addition to the classical action, by
\aln{
\Gamma[G]  &{=F[G]=-\log Z} \nonumber\\
&=  -\frac{1}{2} \tr \log G + \frac{1}{2} \tr(G_0^{-1} G-1) +\Gamma_2[G],
\label{2PIEA}
}
where $\Gamma_2$ is {$(-1)$ times} a collection of {connected} 2PI {bubble} diagrams, {denoted by $\Phi$ in some literature,} in which all propagators are  the renormalized ones $G$. 
{The} 1PI effective action is given by solving the gap equation
\aln{
 \frac{\delta \Gamma[G]}{\delta G} = - \frac{1}{2} G^{-1} + \left(\frac{1}{2} G_0^{-1} +  \frac{\delta \Gamma_2[G]}{\delta G} \right) =0
 \label{gapequation}
}
and substituting $G$ into $\Gamma$.
From the first logarithmic term, it is straightforward to see that 
we have 
\aln{
S^{\text{2PI}}_{\text{1-loop}} =- \frac{V_{d-1} }{12} 
 \int^{1/\epsilon} \frac{d^{d-1}k_{\parallel}} {(2 \pi)^{d-1}}  
  \log \left[ \tilde{G}^{-1}(\bm{0}; k_\parallel) \epsilon^2 \right],
\label{EE-1loop2PI}
}
where $\tilde{G}(\bm{k}; k_\parallel)$ is a Fourier transform of the
renormalized Green function, $G(\bm{x}; x_\parallel)$.
Other contributions to EE follow
 from the second term in Eq. (\ref{2PIEA}) and 2PI diagrams $\Gamma_2$. 
On each plaquette, a flux of twist $m_i$ is assigned. 
Let us first focus on contributions to EE from twisting one of the renormalized propagators in Feynman diagrams. 
By taking a variation with respect to a propagator $G$ and multiply a twisted propagator, 
these contributions are given by
\aln{
& \sum_{n \neq 0} 
\int d^{d+1}x d^{d+1}y
\left (\frac{1}{2} G_0^{-1} +  \frac{\delta \Gamma_2[G]}{\delta G} \right)_{xy} G(\hat{g}^{n}x, y)
\nonumber \\
&= \sum_{n \neq 0} 
\int d^{d+1}x d^{d+1}y
\left (\frac{1}{2} G^{-1} \right)_{yx} G(\hat{g}^{n}x, y).
\label{trivialEE}
}
It is nothing but a twist of $\tr (G^{-1} G) /2$, and gives a trivial result.
Thus, only the logarithmic term of Eq. (\ref{EE-1loop2PI}) provides the EE associated with 
a single twist of a propagator in the 2PI formalism: within the Gaussian approximation, this is a
general result
and consistent with the leading order of perturbative calculations in \cite{Hertzberg:2012mn,Chen:2020ild}.

Among other contributions to EE, the figure-eight diagram in $\Gamma_2$ 
gives the same form of EE as Eq.(\ref{EE-2loop-vertex}),
with $G_0$ replaced by $G$. The next nontrivial contribution to EE comes from the 3-loop diagram in Fig.\ref{Fig5}.
\begin{figure}[h]
\begin{tabular}{c}
	\hspace*{-0.05\linewidth}
	\begin{minipage}{0.5\hsize}
		\centering
		\includegraphics[width=\linewidth]{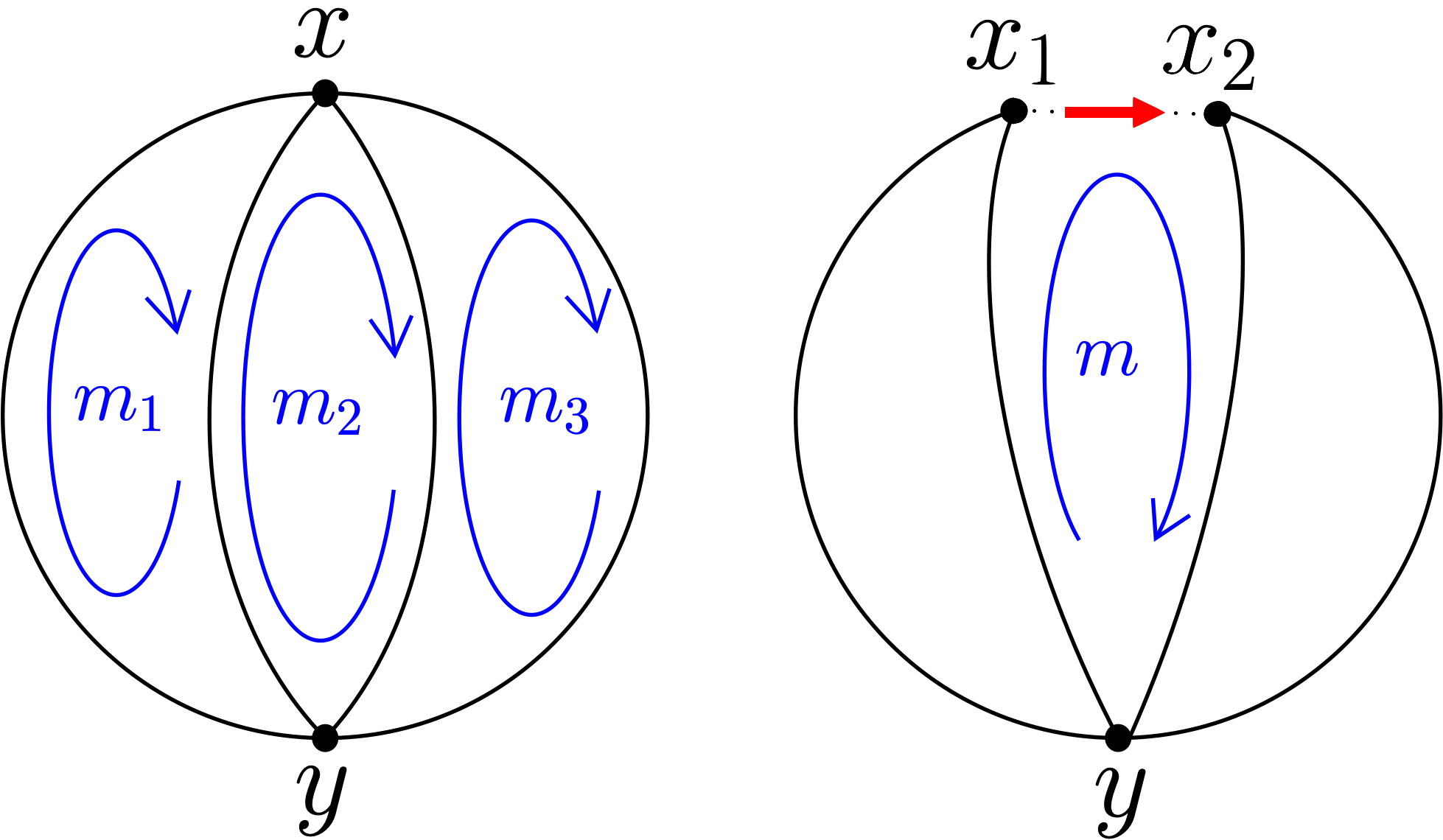}
	\end{minipage}
	\hspace*{0.03\linewidth}
	\begin{minipage}{0.5\hsize}
		\centering
		\includegraphics[width=\linewidth]{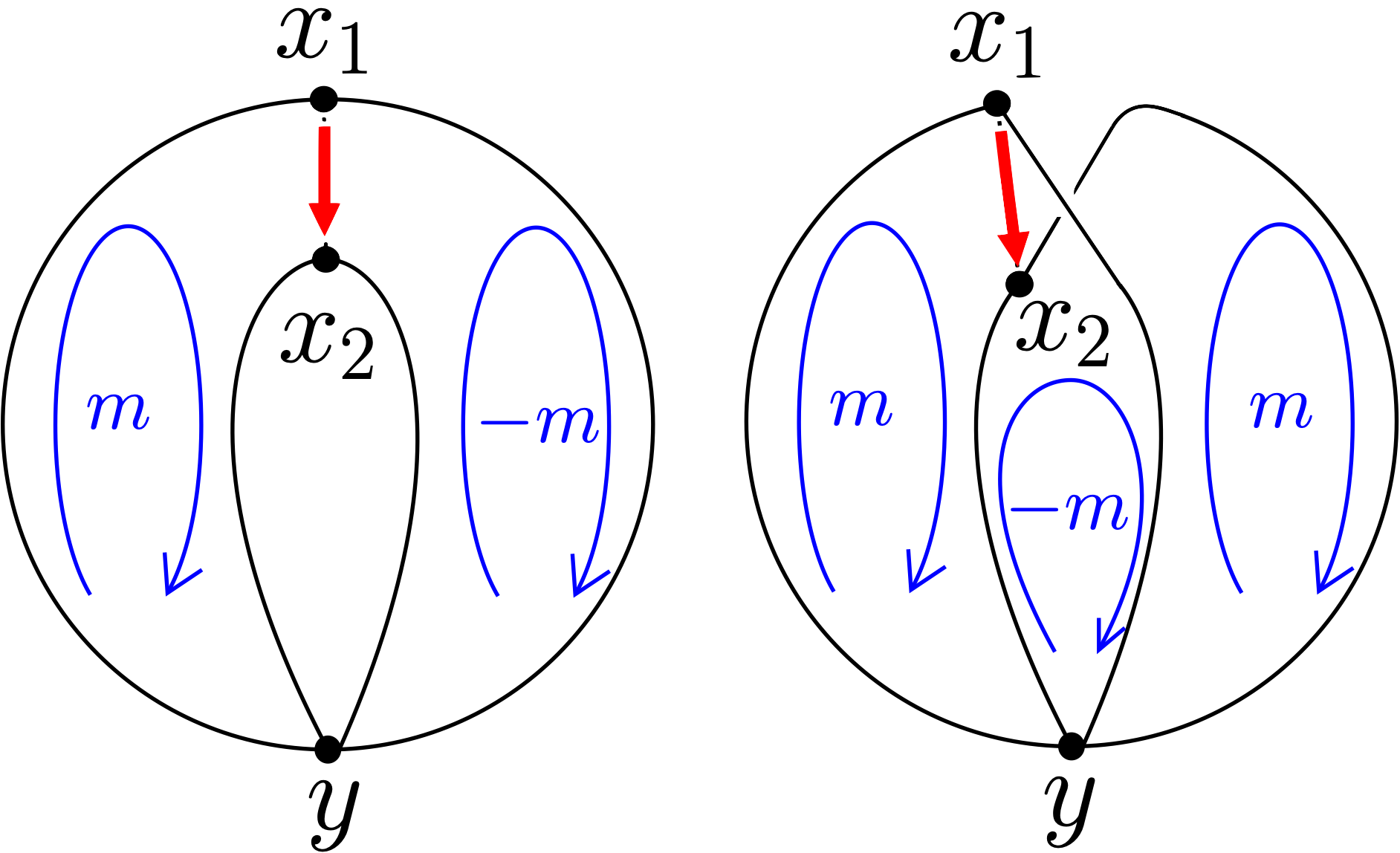}
	\end{minipage}
\end{tabular}
\caption{
A 3-loop diagram  with twists $(m_1, m_2, m_3)$ {(leftmost)}.
A particular configuration $(0,m,0)$ corresponds to twisting a vertex, as well 
as {$(m,0,-m)$ and $(m,-m,m)$} {(three diagrams on the right)}. {These three diagrams are equivalent although they seem different. All of these three diagrams are a single twist of the delta function from $x_1$ to $x_2$.}
}
\label{Fig5}
\end{figure}
{Single twists of propagators,}
as shown in Eq.(\ref{trivialEE}),
vanish in the 2PI formalism by using the gap equation. 
Some other configurations of twists are interpreted as twists of vertices. 
They are given by $(0, m, 0)$ or {$(m,0,-m)$ or $(m,-m,m)$} in Fig.\ref{Fig5}. 
These configurations are regarded as $s, t, u$-channel for twisting the 4-point vertices. {All of them give the same vertex correction.} 
Each configuration of the twists can be interpreted as either
twist of the
upper or lower vertex (but not both). The corresponding EE is given by
\aln{
& S_{\text{3-loop}}^{\text{vertex}} \nonumber\\
&
{
	=-\pdv{M}\Bigg[-3\times \frac{3\lambda^2}{{4}}\int \dd[d+1]{x_1} \dd[d+1]{x_2} \dd[d+1]{y} 
}
\nonumber \\
& \hspace{10mm} 
{
	\times (G(x_1,y))^2 (G(x_2,y))^2 \sum_{m=0}^{M-1} \delta^{d+1} (\hat{g}^{-m} x_1 - x_2) \Bigg]\Bigg\vert_{M\rightarrow 1}
}
\nonumber\\
&= {\frac{3}{2}} V_{d-1} \lambda^2 \int d^2 \bm{x} d^2 \bm{y}
d^{d-1}r_\parallel 
 \nonumber \\
& \hspace{10mm} \times ( G (\bm{x}-\bm{y} ; r_\parallel ) )^2
( G (\bm{x}+\bm{y}; r_\parallel) )^2
  \nonumber \\
&= V_{d-1} \lambda \int d^{d+1}r \  ( G (r) )^2  
\left[  \frac{3  \lambda }{{8}} \int d^2 \bm{X} 
( G (\bm{X}; r_\parallel) )^2 \right] .
}
Comparing it to Eq.(\ref{EE-2loop-vertex}), 
the delta function $\delta^{d-1}(r_\parallel)$, 
which follows twisting the bare 4-point function, 
is replaced by the square bracket in $S_{\text{3-loop}}^{\text{vertex}}$.
The integral including two Green functions
 might be interpreted as twisting a renormalized 4-point vertex function $V_4(x_1, x_2, x_3, x_4)$ at 1-loop, 
as inferred from the right figure of Fig.\ref{Fig5}.
To systematically formulate twisting of higher point functions, 
we need to evaluate, e.g., $\sum_{m \neq 0} V_4(\hat{g}^m x_1, \hat{g}^m x_2, x_3, x_4)$. 
We would like to come back to this issue in future investigations.

\section{V. Conclusions and Discussions}
We have calculated entanglement entropy (EE) of a scalar field theory with $\phi^4$-interactions
in the 2PI formalism
and showed that EE has two different kinds of contributions, one from propagators and
another from vertices. The contributions from propagators are written in terms of
renormalized 2-point Green functions.
On the other hand, those from vertices reflect the non-Gaussian nature of the vacuum wave function. 
The calculations are performed by interpreting
the free energy in terms of 
$\mathbb{Z}_M$ ($M \rightarrow 1)$ gauge theory on Feynman diagrams;
$\mathbb{Z}_M$ fluxes are assigned on each plaquette.
Special configurations of fluxes give the above two contributions. 
Due to the $\mathbb{Z}_M$ twisting, center coordinates of propagators or
positions of vertices are pinned at the origin of the $\mathbb{Z}_M$ orbifold
so that the area law of EE appears. 

There are many issues to be solved. 
We have perturbatively
calculated contributions from 4-point vertices up to 3-loops in the 2PI formalism.
In contrast to the clear understanding of contributions from propagators,
it is  difficult to systematically understand vertex contributions
in terms of fully renormalized 4- (and higher) point functions. 
Besides twisting a single propagator or a vertex, there are many other
configurations of twists. 
The next simple configuration of twists will be twisting two separate propagators.  
We expect that it gives less dominant contributions to EE 
because two positions are simultaneously pinned at the origin due to the twisting, 
and the integration will be largely constrained in Feynman diagram integrals.
{This expectation is} 
also plausible since, if two twists can be independently summed,
each summation gives an $(M^2-1)$ factor and in total $(M^2-1)^2$. Then it does not contribute to EE.
In general, they cannot be independent, but if we can introduce “distance" between twists, 
we could estimate their degrees of contributions to EE. 
For this,  we need 
{a deeper understanding of} $\mathbb{Z}_M$ gauge theory on Feynman diagrams. 
 
{Finally, we comment on the analytical continuation of $M$ to $M \sim 1$. 
The basic assumption of the orbifold method to calculate EE is 
 an analytical continuation from an integer $M$ to a real number.
 It is justified if there are no contributions to EE that vanish at integer $M$s. 
Then, the EE can be calculated by summing all the configurations of 
fluxes of twists on each Feynman diagram. 
In comparison, the heat kernel calculation of EE by Hertzberg \cite{Hertzberg:2012mn} uses a propagator on a cone with an arbitrary deficit angle and no other modifications are made besides propagators.
Our study indicates that in addition to the propagators, vertex functions also need to be modified on a cone. 
It is also interesting to see if some contributions to EE vanish for $1/M$ deficit
angle corresponding to the orbifold case. This will give a justification (or a falsification) for our basic assumption of the
analytical continuation. 
}

\section*{Acknowledgements}
We thank Yoshiki Sato, Sotaro Sugishita, Takao Suyama and Tadashi Takayanagi for 
valuable comments. 
We are supported in part by the Grant-in-Aid for Scientific research, No. 18H03708 (S.I.), No. 16H06490 (S.I.),  No. 20J00079 (K.S.)
and {SOKENDAI}. 


\bibliographystyle{apsrev4-1}
\bibliography{EE}
\end{document}